\documentclass[useAMS,usenatbib,usegraphicx]{mn2e}

\usepackage{amssymb}

\newcommand{\trace}{\mathop{\rm Tr}\nolimits}
\newcommand{\Id}{\mathbfss{I}}
\newcommand{\expect}{\mathbb{E}}
\newcommand{\disp}{\mathbb{D}}
\newcommand{\const}{\mathop{\rm const}\nolimits}
\newcommand{\sh}{\mathop{\rm sh}\nolimits}
\newcommand{\NIF}{{N_{\rm ind}}}
\newcommand{\FAP}{{\rm FAP}}

\title[Assessing statistical significance of periodogram peaks]%
{Assessing statistical significance of periodogram peaks}
\author[R.V.~Baluev]%
{R.V.~Baluev$^{1}$ \thanks{E-mail: roman@astro.spbu.ru}\\
$^{1}$Sobolev Astronomical Institute, Saint Petersburg State University,\\
Universitetskij prospekt 28, Petrodvorets, Saint Petersburg 198504, Russia}
\begin{document}

\date{  Accepted 2007 November --.
        Received 2007 October 16;
in original form 2007 June 13}

\pagerange{\pageref{firstpage}--\pageref{lastpage}} \pubyear{2007}

\maketitle

\label{firstpage}

\begin{abstract}
The least-squares (or Lomb-Scargle) periodogram is a powerful tool which is used routinely
in many branches of astronomy to search for periodicities in observational data. The
problem of assessing statistical significance of candidate periodicities for different
periodograms is considered. Based on results in extreme value theory, improved analytic
estimations of false alarm probabilities are given. They include an upper limit to the
false alarm probability (or a lower limit to the significance). These estimations are
tested numerically in order to establish regions of their practical applicability.
\end{abstract}


\section{Introduction}
Analysing astronomical time series, one often has to choose between at least two
hypotheses, a base one $\mathcal H$ and an alternative one $\mathcal K$, based on the
existing data array. In the signal detection problem, one should check whether the
observations are consistent with some base model or they contain an extra deterministic
signal. Under presence of random errors, such problem can be solved in a probabilistic
sense only. We are never protected from mistakes of two kinds. They are the false
retraction of $\mathcal H$ (the `false alarm') and the false non-retraction of $\mathcal
H$ (the false non-detection). False alarms are generally believed to be more dangerous,
hence the problem of estimation of the false alarm probability (hereafter $\FAP$)
associated with a candidate signal is very important. Given some small critical value
$\FAP_*$ (between $10^{-3}$ and $0.1$ usually), we could claim that the candidate signal
is statistically significant (if its $\FAP<\FAP_*$) or is not ($\FAP>\FAP_*$).

For the \citet{Lomb76}~-- \citet{Scargle82} periodogram (hereafter also L--S), the base
hypothesis is that the observations are pure zero-mean uncorrelated and Gaussian errors
(also called the white Gaussian noise). The alternative one is that a sinuous harmonic is
also present. Every single value of the L--S periodogram represents a test statistic for
the corresponding problem of hypotheses testing. In routine practical cases, however, the
period of a possible signal is not known a priory and we have to scan many periodogram
values within a wide frequency range. In this case, the $\FAP$ is provided by the
probability distribution of the maximum periodogram value under the base hypothesis (i.e.,
without signal in the data). Existing methods of calculating this distribution for a
continuous frequency range require time-consuming Monte-Carlo simulations. The aim of the
present paper is to propose analytic approximations which could allow to avoid Monte-Carlo
simulations (at least in many practical cases). Such approximations of the distribution of
the maximum have already been constructed by mathematicians specializing in the field of
extreme values of random processes. In the Section~\ref{sec_Stat}, these results are
adapted for and extended to the specific features of the periodogram analysis of
astronomical time series. In the Section~\ref{sec_Simul}, numerical simulations are used
to explore the quality of the analytic results and to show regions of their practical
applicability.

\section{General formulations}
\label{sec_Intro}
Let us recover the principles of the periodogram analysis in a somewhat more general
formulation than usually.%
\footnote{Several mathematical notations, used in the present paper, are described in the
Appendix~\ref{sec_not}.}%

Let $x_1,x_2\ldots x_N$ be observations made at $N$ epochs $t_1,t_2,\ldots t_N$. The
errors of $x_i$ are assumed to be independent and Gaussian with standard deviations
$\sigma_i$. Each value of the periodogram can be recovered as a test statistic that allows
to conclude, how likely is the hypothesis that the data contain a signal of a given
frequency $f$. Mathematically we should check, whether the observations are fitted well by
some base model having only $d_{\mathcal H}$ free parameters $\btheta_{\mathcal H}$, or
they require an enlarged model of $d_{\mathcal K}$ parameters $\btheta_{\mathcal K} =
\{\btheta_{\mathcal H}, \btheta\}$ with $d = d_{\mathcal K} - d_{\mathcal H}$ parameters
$\btheta$ of an extra periodic signal. We will assume that for any fixed frequency both
models are linear and construct them by means of $d_{\mathcal H}$ and $d_{\mathcal K}$
base functions forming vectors $\bvarphi_{\mathcal H}(t)$ and $\bvarphi_{\mathcal K}(t,f)
= \{\bvarphi_{\mathcal H}(t), \bvarphi(t,f)\}$. Thus the base fit model is $\mu_{\mathcal
H}(t, \btheta_{\mathcal H}) = \btheta_{\mathcal H} \cdot \bvarphi_{\mathcal H}(t)$, the
model of the signal is $\mu(t, \btheta, f) = \btheta \cdot \bvarphi(t,f)$ and the complete
fit model is $\mu_{\mathcal K}(t, \btheta_{\mathcal K}, f) = \btheta_{\mathcal K} \cdot
\bvarphi_{\mathcal K}(t,f) = \mu_{\mathcal H}(t, \btheta_{\mathcal H}) + \mu(t, \btheta,
f)$. We wish to test, whether the hypothesis $\mathcal H: \btheta=0$ should be rejected in
favour of the alternative $\mathcal K(f): \btheta\neq 0$.

For the L--S periodogram $d_{\mathcal H}=0$, $d=2$, and the signal model is given by a
harmonic function $\theta_1 \cos\omega t + \theta_2 \sin\omega t$ (here $\omega=2\pi f$).
\citet{SchwCzerny98a,SchwCzerny98b} considered cases with $d_{\mathcal H}=0$ and arbitrary
$d$. \citet{FerrazMello81} put $d_{\mathcal H}=1$ and added a floating constant term to
the harmonic model with $d=2$, whereas \citet{Cumming99} accounted also for possible
linear trend ($d_{\mathcal H}=2$).

An optimal statistical test, solving such problem in general, is developed rather well
\citep[chapter~7]{Lehman}. At first, one should compute the minima (by $\btheta_{\mathcal
H, \mathcal K}$) of the function $\chi^2 = \left\langle (x - \mu_{\mathcal K})^2
\right\rangle$ under hypotheses $\mathcal H$ and $\mathcal K(f)$. This may be done by
means of any accessible linear least-squares algorithm \citep[see
also][]{SchwCzerny98a,SchwCzerny98b}. If $\sigma_i$ are known precisely, both minima
$\chi_{\mathcal H}^2$ and $\chi_{\mathcal K}^2(f)$ can be computed and the least-squares
periodogram may be defined as an advance in $\chi^2$ provided by the transition from
$\mathcal H$ to $\mathcal K(f)$:
\begin{equation}
 z(f) = \left[ \chi_{\mathcal H}^2 - \chi_{\mathcal K}^2(f) \right]/2.
\label{PeriodogramDef}
\end{equation}
The error variances are often not known precisely and we have to estimate them from the
time series, explicitly or implicitly. It is usually assumed that $\sigma_i = \kappa
\sigma_{{\rm mes}, i}$, where the `measured' uncertainties $\sigma_{{\rm mes}, i}$
determine the weighting pattern of the time series, whereas the coefficient $\kappa$ is
unconstrained. In this case, only the ratio $\chi_{\mathcal H}^2/\chi_{\mathcal K}^2$ can
be computed exactly, and the periodogram~(\ref{PeriodogramDef}) has to be modified. We
will consider the following modified periodograms:
\begin{eqnarray}
 z_1(f) = N_{\mathcal H} \frac{\chi_{\mathcal H}^2-\chi_{\mathcal K}^2(f)}{2\chi_{\mathcal H}^2}, \quad
 z_2(f) = N_{\mathcal K} \frac{\chi_{\mathcal H}^2-\chi_{\mathcal K}^2(f)}{2\chi_{\mathcal K}^2(f)}, \nonumber\\
 z_3(f) = \frac{N_{\mathcal K}}{2} \ln \frac{\chi_{\mathcal H}^2}{\chi_{\mathcal K}^2(f)}.
\label{PeriodogramModDef}
\end{eqnarray}
Here, $N_{\mathcal H}=N-d_{\mathcal H}$ and $N_{\mathcal K}=N-d_{\mathcal K}$ are the
numbers of degrees of freedom in $\chi_{\mathcal H}^2$ and $\chi_{\mathcal K}^2$,
correspondingly. The periodograms $z_1(f)$ and $z_2(f)$ are the well-known normalizations
of $z(f)$ by variances of residuals under the respective hypotheses. All
periodograms~(\ref{PeriodogramModDef}) are entirely equivalent because they are unique
functions of each other:
\begin{eqnarray}
 2 z_1/N_{\mathcal H} = 1 - e^{-2z_3/N_{\mathcal K}}, \quad
 2 z_2/N_{\mathcal K} = e^{2z_3/N_{\mathcal K}} - 1, \nonumber\\
 (1-2z_1/N_{\mathcal H})(1+2z_2/N_{\mathcal K})=1.
\label{feeds}
\end{eqnarray}

\section{False alarm probability}
\label{sec_Stat}
Let us pick any of the periodograms introduced above, and denote it as $Z(f)$. If the
frequency of a possible signal was known, the false alarm probability could be retrieved
as $\FAP_{\rm single} = 1 - P_{\rm single}(Z)$, where $P_{\rm single}(Z)$ is the
cumulative distribution function of $Z(f)$ (taken under the base hypothesis). Under the
hypothesis $\mathcal H$, the statistic $2z$ follows a $\chi^2$-distribution with $d$
degrees of freedom, $2z_2/d$ obeys a Fisher-Snedecor $F$-distribution with $d$ and
$N_{\mathcal K}$ degrees of freedom, and $2z_1/N_{\mathcal H}$ obeys a beta distribution
with the same numbers of degrees of freedom \citep[\S7.1]{Lehman}. Using
relations~(\ref{feeds}), the distribution function of $z_3$ can be derived easily. The
corresponding expressions of false alarm probability for $d=2$ are given in
Table~\ref{tab_FAP}. Note that the third modified periodogram obeys exactly the same
distribution as the basic one, if $d=2$.

Now let us assume that we scan all frequencies from the interval $[0,f_{\rm max}]$ and
look for the maximum value $Z_{\rm max} = \max_{[0,f_{\rm max}]} Z(f)$. Then the false
alarm probability, associated with this maximum, is $\FAP_{\rm max} = 1 - P_{\rm
max}(Z_{\rm max},f_{\rm max})$, where $P_{\rm max}(Z_{\rm max},f_{\rm max})$ denotes the
cumulative distribution function of $Z_{\rm max}$ (under the base hypothesis). Precise
expression for the latter distribution is not known even for equally spaced time series.
It is always possible to use Monte-Carlo simulations to obtain this function, but this way
is very time-consuming, especially for the most important region of low false alarm
probabilities (high significances). The function $P_{\rm max}(Z,f_{\rm max})$ is often
computed~\citep{SchwCzerny98a,SchwCzerny98b} as
\begin{equation}
 P_{\rm max}(Z,f_{\rm max}) \approx P_{\rm single}(Z)^{\NIF(f_{\rm max})},
\label{mult-trial}
\end{equation}
where $\NIF(f_{\rm max})$ is an effective `number of independent frequencies' found within
$[0,f_{\rm max}]$. There is no general analytic expression for the quantity $\NIF$, but it
is often suggested to use a short Monte-Carlo simulation to assess it and then
extrapolate~(\ref{mult-trial}) to low $\FAP$ \citep{Cumming04,HorneBal86}. However, the
multiple-trial formula~(\ref{mult-trial}) is only heuristic and is not necessarily precise
even for equally spaced observations which don't produce significant aliasing.

A better estimation of $P_{\rm max}(Z,f_{\rm max})$ may be obtained using the theory of
stochastic processes. The theory of extremes of random processes is developed in
mathematical literature rather deeply. For our aims, it is worth to mention the series of
works by \citet{Davies77,Davies87,Davies02}. This author considered (in rather general
formulations) extreme value distributions for $\chi^2$, $F$, and beta random processes
that may include our periodograms $z$ and $z_{1,2}$ as special cases. The main result of
these works is an analytic lower limit to the corresponding extreme value distributions.
This result is potentially very useful for astronomical applications, because it yields
directly an upper limit to the false alarm probability and a lower limit to the
significance of a candidate periodicity. However, the formulae published in the cited
papers are not yet ready for usage and require some adaptation to specific applications.
Moreover, these results can be improved to obtain not only an upper limit, but an uniform
approximation to the false alarm probability, that would be good for low spectral leakage
at least.

A brief description of these results, adapted for the uneven time series analysis, along
with detailes of my extensions, is given in the Appendix~\ref{sec_RiceSeries}. Summarizing
them, the `Davies bound' may be written down as
\begin{equation}
 \FAP_{\rm max}(Z,f_{\rm max}) \leq \FAP_{\rm single}(Z) + \tau(Z,f_{\rm max}).
\label{simFAP}
\end{equation}
The function $\tau$ will be specified below. If the aliasing effects may be neglected
within the frequency band being scanned\footnote{This means that the spectral window of
the time series has no significant peaks in the doubled frequency band $[0,2f_{\rm max}]$,
except for the main one at $f=0$.}, and if also $f_{\rm max}$ is large enough, then
\begin{eqnarray}
 P_{\rm max}(Z,f_{\rm max}) \approx P_{\rm single}(Z)\, e^{-\tau(Z,f_{\rm max})}.
\label{expo}
\end{eqnarray}
The right hand side in~(\ref{simFAP}) should approach the false alarm probability more
closely for large $Z$ (even the asymptotic equality under $Z\to\infty$ is expected, but
not proved strictly yet). In general, the quantity $\tau(Z,f_{\rm max})$ looks like
\begin{equation}
 \tau = \left(\frac{z}{\pi}\right)^{\frac{d-1}{2}} \frac{e^{-z}}{2\pi} A(f_{\rm max})
\label{tau_z}
\end{equation}
for the basic least-squares periodogram~(\ref{PeriodogramDef}) and like
\begin{equation}
 \tau = \frac{\gamma}{2\pi} A(f_{\rm max}) \times
 \left\{\begin{array}{@{}l}
  \left( \frac{2z_1}{\pi N_{\mathcal H}} \right)^{\frac{d-1}{2}}
   \left( 1-\frac{2z_1}{N_{\mathcal H}} \right)^{\frac{N_{\mathcal K}-1}{2}}, \\
  \left( \frac{2z_2}{\pi N_{\mathcal K}} \right)^{\frac{d-1}{2}}
   \left( 1+\frac{2z_2}{N_{\mathcal K}} \right)^{-\frac{N_{\mathcal H}}{2}+1}, \\
  \left( \frac{2}{\pi}\sh\frac{z_3}{N_{\mathcal K}} \right)^{\frac{d-1}{2}}
   e^{-z_3 \left( 1+ \frac{d-3}{2N_{\mathcal K}} \right)}.
 \end{array}\right.
\label{tau_z123}
\end{equation}
for the modified periodograms~(\ref{PeriodogramModDef}). Here the coefficient $\gamma=
\Gamma\left(\frac{N_{\mathcal H}}{2}\right)\left/\Gamma\left(\frac{N_{\mathcal
K}+1}{2}\right)\right.$. Note that the asymptotic $(2/N_{\mathcal H,\mathcal
K})^{(d-1)/2}\gamma\to 1$ holds true for $N\to \infty$. The factor $A(f_{\rm max})$
depends on the bases $\bvarphi$ and $\bvarphi_{\mathcal H}$, on the time series sampling
and on the weighting pattern. Unfortunately, the general form of $A(f_{\rm max})$,
obtained in the Appendix~\ref{sec_RiceSeries}, is not simple. For now, let us restrict
ourselves to the L--S periodograms and neglect by aliasing effects. In the next section we
will show that such approximation for $A(f_{\rm max})$ works well even for strong
aliasing. Of course, anyone is welcome to calculate $A(f_{\rm max})$ numerically from the
formulae given in the Appendix: such work is still much less computationaly expensive than
Monte-Carlo simulation of $P_{\rm max}(Z,f_{\rm max})$. The practical quality of the
expressions~(\ref{simFAP},\ref{expo}) will be also explored numerically in the next
section.

To derive $A(f_{\rm max})$ from the formula~(\ref{coeffA}), we should calculate the
eigenvalues of the matrix $\mathbfss M$, which is defined by the group of
equalities~(\ref{defM}). To perform this, we have to concretize the functions
$\bvarphi(t,f)$. For the usual L--S periodogram the harmonic base
\begin{equation}
 \bvarphi(t,f) = \{ \cos\omega t, \sin\omega t \} \quad (\omega=2\pi f)
\label{sincos}
\end{equation}
produces the matrices
\begin{eqnarray}
\mathbfss Q &=& \frac{1}{2} \left(\begin{array}{cc}
    1 + \overline{\cos 2\omega t} & \overline{\sin 2\omega t}     \\
    \overline{\sin 2\omega t}     & 1 - \overline{\cos 2\omega t}
   \end{array}\right), \nonumber\\
\mathbfss S &=& \pi \left(\begin{array}{cc}
    -\overline{t \sin 2\omega t} & \bar t + \overline{t \cos 2\omega t} \\
    -\bar t + \overline{t \cos 2\omega t} & \overline{t \sin 2\omega t}
   \end{array}\right), \nonumber\\
\mathbfss R &=& 2\pi^2 \left(\begin{array}{cc}
    \overline{t^2} - \overline{t^2 \cos 2\omega t} & - \overline{t^2 \sin 2\omega t} \\
    - \overline{t^2 \sin 2\omega t} & \overline{t^2} + \overline{t^2 \cos 2\omega t}
   \end{array}\right), \nonumber\\
\mathbfss M &=& \mathbfss Q^{-1}(\mathbfss R - \mathbfss S^T \mathbfss Q^{-1} \mathbfss S).
\label{harmQSR}
\end{eqnarray}
If we consider the alias-free case, the terms in~(\ref{harmQSR}) containing sine and
cosine functions of frequencies $2f\leq 2f_{\rm max}$ are averaged out. Under this
approximation $\mathbfss M \approx 4\pi^2 \disp t\, \Id$, where $\disp t = \overline{t^2}
- \bar t^2$ is the weighted variance of the observational epochs. Then both eigenvalues
required are equal to the constant $4\pi^2 \disp t$ and $A(f_{\rm max})\approx 2\pi^{3/2}
W$, where $W = f_{\rm max}T_{\rm eff}$ is a rescaled frequency bandwidth and $T_{\rm eff}
= \sqrt{4\pi \disp t}$ is an effective time series length. If $t_i$ are spanned uniformly
and all $\sigma_i$ are equal, then $T_{\rm eff}$ almost coincides with an actual time
series span. Table~\ref{tab_FAP} contains the alias-free approximations of $\tau(Z,f_{\rm
max})$ for all L--S periodograms considered. One may use these expressions and the
ones~(\ref{simFAP},\ref{expo}) to write down the corresponding alias-free approximation of
$P_{\rm max}(Z,f_{\rm max})$ and its Davies bound. Routinely we deal with rather large
values of $z$ and $W$. In this case either the factor $P_{\rm single}(Z)$ in~(\ref{expo})
or the term $\FAP_{\rm single}(Z)$ in~(\ref{simFAP}) may be safely neglected. For
instance, for the usual L--S periodogram
\begin{equation}
 P_{\rm max}(z,f_{\rm max}) \approx (1-e^{-z})\, e^{-We^{-z}\sqrt{z}}
                            \approx e^{-We^{-z}\sqrt{z}}.
\label{MaximaDistr}
\end{equation}
Such alias-free approximations are valid if $f_{\rm max}$ is well resolved ($W \gtrsim 1$)
and if the spectral leakage is low. Only the latter assumption is practically significant.
If one worries about strong spectral leakage, the approximate inequality
\begin{equation}
 \FAP_{\rm max}(z,f_{\rm max}) \lessapprox e^{-z} + W e^{-z} \sqrt{z} \approx W e^{-z} \sqrt{z},
\label{MaximaDistrIneq}
\end{equation}
holds true for the basic L--S periodogram. The
relations~(\ref{MaximaDistr},\ref{MaximaDistrIneq}) are equally valid if the base model is
not empty, but includes a low-order polynomial drift and/or several harmonics of fixed
frequencies that may be considered as independent on any frequency within the range being
scanned.

\begin{table}
\caption{False alarm probabilities for the Lomb--Scargle periodogram and its modifications
($d=2$).}
\label{tab_FAP}
\begin{tabular}{@{}lll}
\hline
$Z(f)$   & ${\rm FAP}_{\rm single}(Z)$ & $\tau(Z,f_{\rm max})$, approximately \\
\hline
$z(f)$   & $e^{-Z}$ & $\phantom{\gamma_{\mathcal K}\,} W e^{-Z}\sqrt{Z}$\\
$z_1(f)$ & $\left(1-\frac{2Z}{N_{\mathcal H}}\right)^{\frac{N_{\mathcal K}}{2}}$ &
     $\gamma_{\mathcal H}\, W \left(1-\frac{2Z}{N_{\mathcal H}}\right)^{\frac{N_{\mathcal K}-1}{2}}\sqrt{Z}$ \\
$z_2(f)$ & $\left(1+\frac{2Z}{N_{\mathcal K}}\right)^{-\frac{N_{\mathcal K}}{2}}$ &
     $\gamma_{\mathcal K}\, W \left(1+\frac{2Z}{N_{\mathcal K}}\right)^{-\frac{N_{\mathcal K}}{2}}\sqrt{Z}$ \\
$z_3(f)$ & $e^{-Z}$ & $\gamma_{\mathcal K}\, W e^{-Z\left(1-\frac{1}{2N_{\mathcal K}}\right)}
     \sqrt{N_{\mathcal K}\sh\frac{Z}{N_{\mathcal K}}}$ \\
\hline
\end{tabular}\\
The factors $\gamma_{\mathcal H,\mathcal K} = \sqrt{\frac{2}{N_{\mathcal H, \mathcal K}}}
\,\Gamma\left(\frac{N_{\mathcal H}}{2}\right) \left/ \Gamma\left(\frac{N_{\mathcal H}
-1}{2}\right)\right.$ may be neglected for $N_{\mathcal H}\geq 10$. If the spectral
leakage is low, $\FAP_{\rm max}\approx \tau(Z,f_{\rm max})$ for realistic values of
parameters (see text).
\end{table}

For large $N$, every modified periodogram $z_{1,2,3}$ obeys approximately the same extreme
value distribution as the basic one. However, this convergence is not uniform in $Z$. It
is easy to derive from~(\ref{tau_z},\ref{tau_z123}) that for the periodograms $z_{1,2}(f)$
an extra condition $Z\ll \sqrt N$ must be satisfied to keep relative errors of $\FAP$ low.
This condition is rarely satisfied in practice. For the periodogram $z_3$, a corresponding
condition $Z\ll N$ is mild and is often satisfied in practical applications. This fact
requires to consider the third modified periodogram more closely. The log-likelihood
function of our Gaussian observations is given by
\begin{equation}
 \ln \mathcal L =  - \chi^2/2 - \sum_{i=1}^{N} \ln \sigma_i + \const.
\end{equation}
As we adopted $\sigma_i = \kappa \sigma_{{\rm mes}, i}$, this expression may be rewritten
as $\ln\mathcal L = - \tilde{\chi}^2/(2 \kappa^2) - N \ln\kappa + \const$, where
$\tilde\chi^2$ does not depend on $\kappa$. Maximizing $\ln \mathcal L$ by $\kappa$ under
the hypotheses $\mathcal K$ and $\mathcal H$ yields that the logarithm of the ratio of the
corresponding likelihood maxima equals to $\frac{N_{\mathcal K}}{N} z_3$.

\section{Numerical simulations}
\label{sec_Simul}
Let us test the analytic results introduced above. For this purpose, we will use
simulations of time series of $N$ quasi\-random data points imitating the white Gaussian
noise. The temporal moments $t_i$ cover a segment of a length $T$. The uncertainties
$\sigma_i$ are equal to each other unless otherwise stated. For every simulation discussed
below, no less than $10^5$ Monte-Carlo trials were generated ($t_i$ and $\sigma_i$ were
fixed during every such simulation, of course). This should provide accuracies of
simulated $\FAP$s about $1\%$ for $\FAP=0.1$ and about $10\%$ for $\FAP=10^{-3}$. The
simulated tale of $\FAP<10^{-3}$ often showed unstable deviations comparable with the
false alarm probability.

\begin{figure}
\includegraphics[width=84mm]{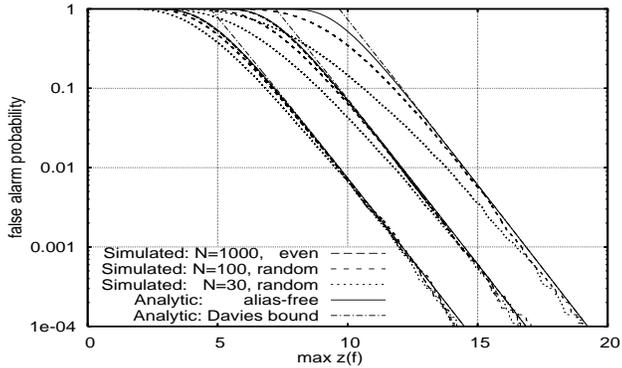}
\caption{Simulated vs. analytic false alarm probability for the L--S periodogram with no
forced data gapping. Simulations for $1000$ evenly, $100$ and $30$ randomly spaced
observations, $10^5$ Monte-Carlo trials, $f_{\rm max} T = 50, 500, 5000$ (bunches from
left to right). In the even cases, the simulated curves almost coincide with their
alias-free approximations. Due to exceedance of the Nyquist frequency, there is no curve
for the even case and $f_{\rm max}T = 5000$. All graphs of analytic expressions are
plotted for $T_{\rm eff}= T$ (this equality holds true within a few per cent for the three
time series used here).}
\label{fig_simDistr}
\end{figure}

\begin{figure}
\includegraphics[width=84mm]{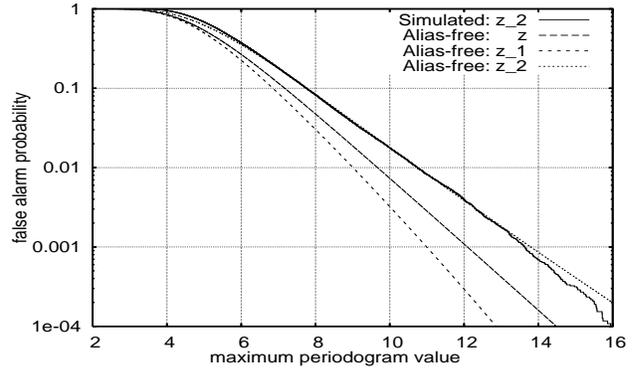}
\caption{Simulated vs. analytic false alarm probability for the modification $z_2$ of
the L--S periodogram: $100$ evenly spaced observations, $10^5$ Monte-Carlo trials, $f_{\rm
max} T = 50$. For comparison, the theoretical distribution curves for the periodograms $z$
and $z_1$ are also plotted (the curve for $z_3$ almost coincides with that for $z$ and is
not shown).}
\label{fig_simDistrz2}
\end{figure}

If the time series consists of large number of equally spaced observations, any aliasing
should be negligible. Indeed, in such a case the Davies bound~(\ref{MaximaDistrIneq})
appears very sharp (for $\FAP<0.1$) and the analytic approximation~(\ref{MaximaDistr})
perfectly follows the simulated distribution
(Figs.~\ref{fig_simDistr},\ref{fig_simDistrz2}). However, even time series don't allow to
search frequencies less than the Nyquist one $f_{\rm Ny} = (N-1)/(2T)$. An uneven time
series allows to access much lower frequencies. However, this access cannot be free of any
charge. Within a wide frequency range ($f_{\rm max} \gtrsim f_{\rm Ny}$), an essential
aliasing is normally present purely due to random fluctuations of observational moments,
even if there is no physical necessity for their gapping. Thus we may expect that for
$W\gtrsim N$ a significant `natural' aliasing should take place. According to the
numerical results shown on Fig.~\ref{fig_simDistr}, the alias-free approximation indeed
becomes significantly less precise when $N$ decreases, but for large $\FAP$ only (say,
larger than a few per cent). Even for $W>100 N$ the loss of precision remains moderate for
practically important values of $\FAP$. We can quite use~(\ref{MaximaDistr}) for practical
calculations even if $W$ is ten times larger than $N$ (or even larger, depending on the
desirable precision).

\begin{figure}
\includegraphics[width=84mm]{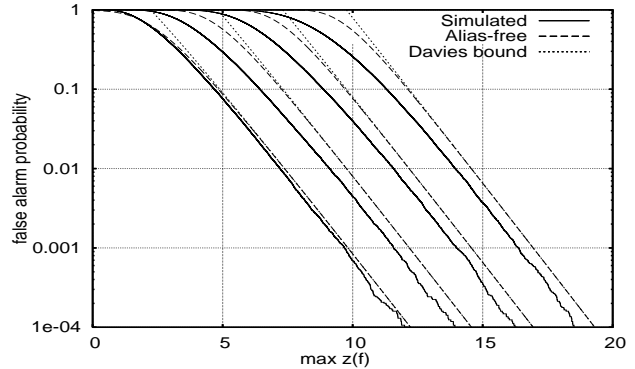}
\caption{Simulated vs. analytic false alarm probability for the L--S periodogram with
forced data gapping. $N=100$ observations were clumped in ten equal groups, each group
spans (randomly) only a fifth fraction of its natural duration. About $1.7\cdot 10^5$
Monte-Carlo trials, $f_{\rm max} T = 5, 50, 500, 5000$ (from left to right).}
\label{fig_simDistrAlias}
\end{figure}

When a `physical' spectral leakage is large, the quality of the alias-free approximation
depends on the frequency range too. If $f_{\rm max}$ does not exceed the Nyquist frequency
of periodic breaking of observations then the interval $[0,f_{\rm max}]$ is free from
aliasing and we may use~(\ref{MaximaDistr}) without significant loss of precision. If the
frequency range increases, the model~(\ref{MaximaDistr}) comes off from the real
distribution and somewhat overestimates the false alarm probability. Such a simulation is
shown in Fig.~\ref{fig_simDistrAlias}. In this example, the frequency of periodic data
breaks corresponds to $f_{\rm max}T = 9$ and the respective Nyquist frequency corresponds
to a half of this value ($f_{\rm max}T = 4.5$).

Although errors of the alias-free model may become practically significant for some
extremal situations, they are not very large and (more important) not fatal. The
significance of a candidate periodicity is underestimated, what does not favour to false
alarms. The aliasing may only decrease the detectability of low-amplitude signals (if
numerical simulations are not used). In this case, the error of the threshold level (i.e.,
the critical level $z_*$, corresponding to a given $\FAP_*$) is more important. Examining
Fig.~\ref{fig_simDistrAlias} yields that the relative shift $\Delta z_* / z_*$ does not
exceed $10\%$ for $\FAP<0.1$. As an amplitude of corresponding signal scales as
$\sqrt{z}$, this turns into only $5\%$ relative error of an amplitude threshold. Remind,
that this $5\%$ offset corresponds to a very strong aliasing. Such spectral leakage takes
place, for instance, for the sequence of observations that are made during $10$ days with
only $4.8$ night hours in a day, or during $10$ years with only $2.4$ observational months
in a year.

\begin{figure}
\includegraphics[width=84mm]{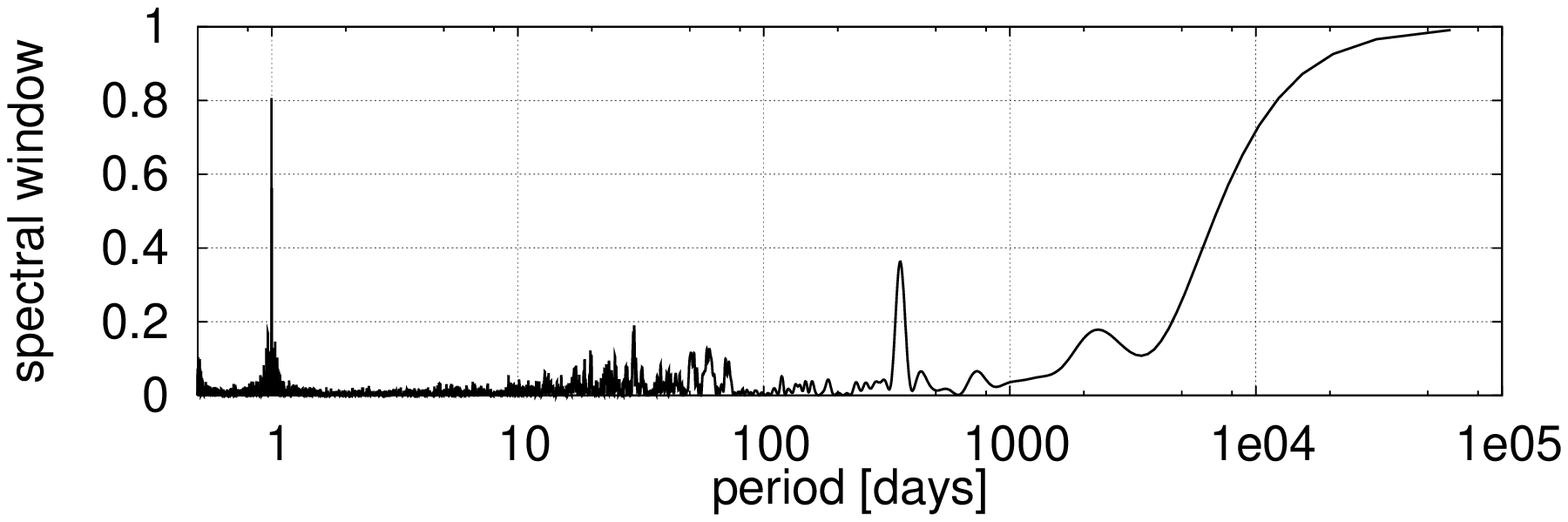}\\
\includegraphics[width=84mm]{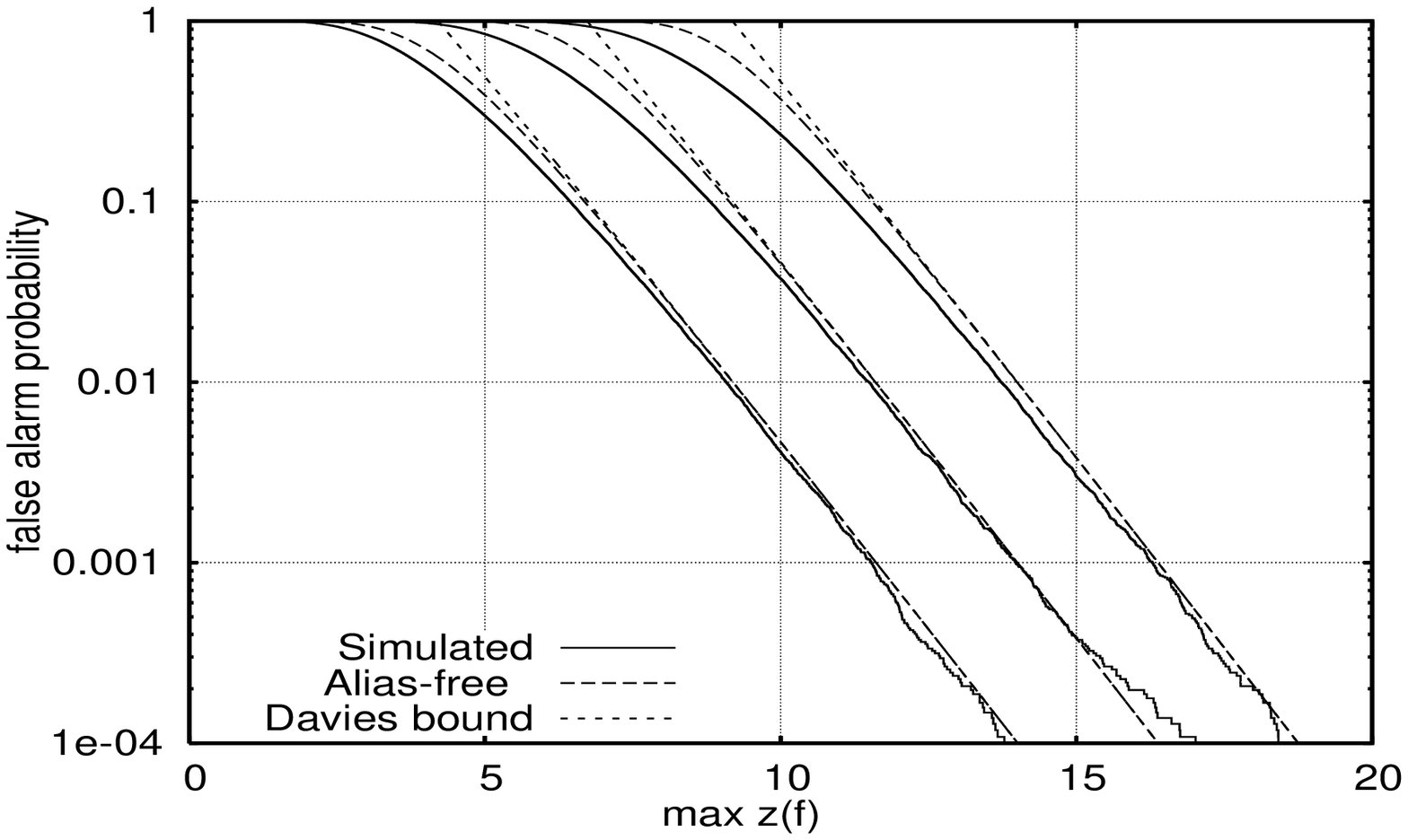}
\caption{Top: spectral window function for ELODIE radial velocities of 51~Peg. Bottom:
Simulated and analytic $\FAP$ for this time series for $10^5$ Monte-Carlo trials, $P_{\rm
min} = 1/f_{\rm max} = 100, 10, 1$~days (from left to right).}
\label{fig_simDistr51Peg}
\end{figure}

Note that the multiple-trial formula~(\ref{mult-trial}) can work well in restricted
regions only. When constructed from a short Monte-Carlo simulation, it can fit well the
centre of the distribution (i.e., large $\FAP$s), but fails to fit low $\FAP$s. This takes
place even for negligible aliasing. The spectral leakage perturbs strongly the
distribution centre but affects weakly its high-significance tail. Hence, any
multiple-trial models constructed from short Monte-Carlo simulations cannot be
extrapolated to the most important region of low false alarm probabilities. Such
extrapolation overestimates statistical significances of candidate periodicities, what
favours to false alarms.

\begin{figure}
\includegraphics[width=84mm]{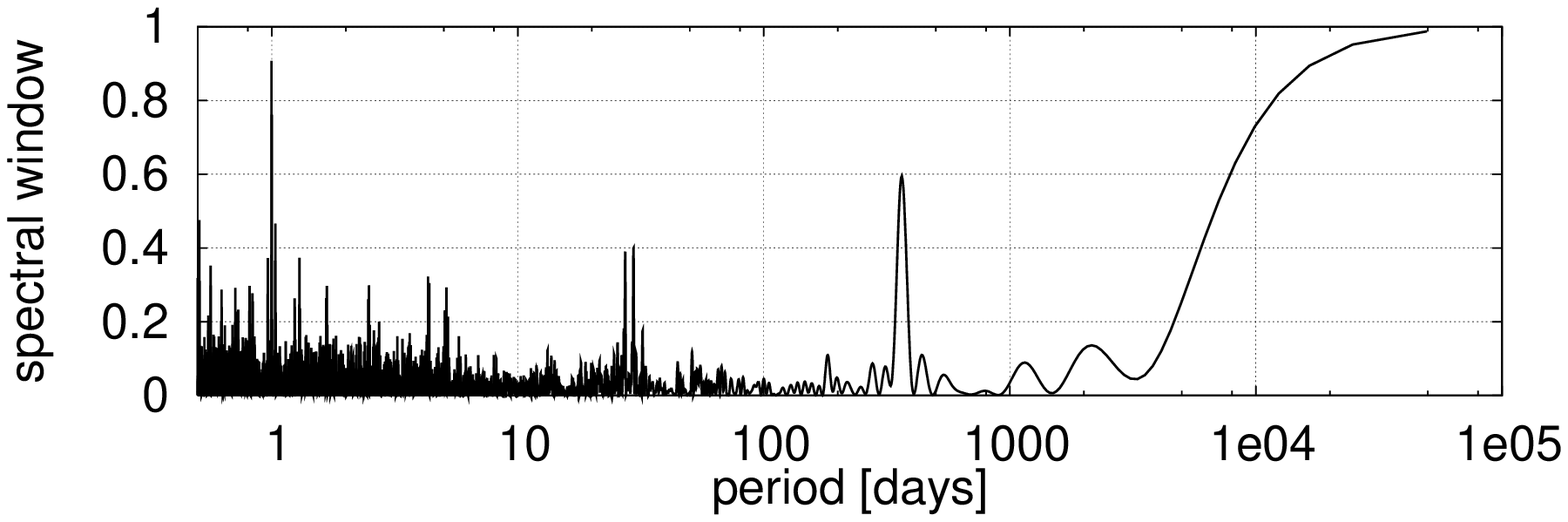}\\
\includegraphics[width=84mm]{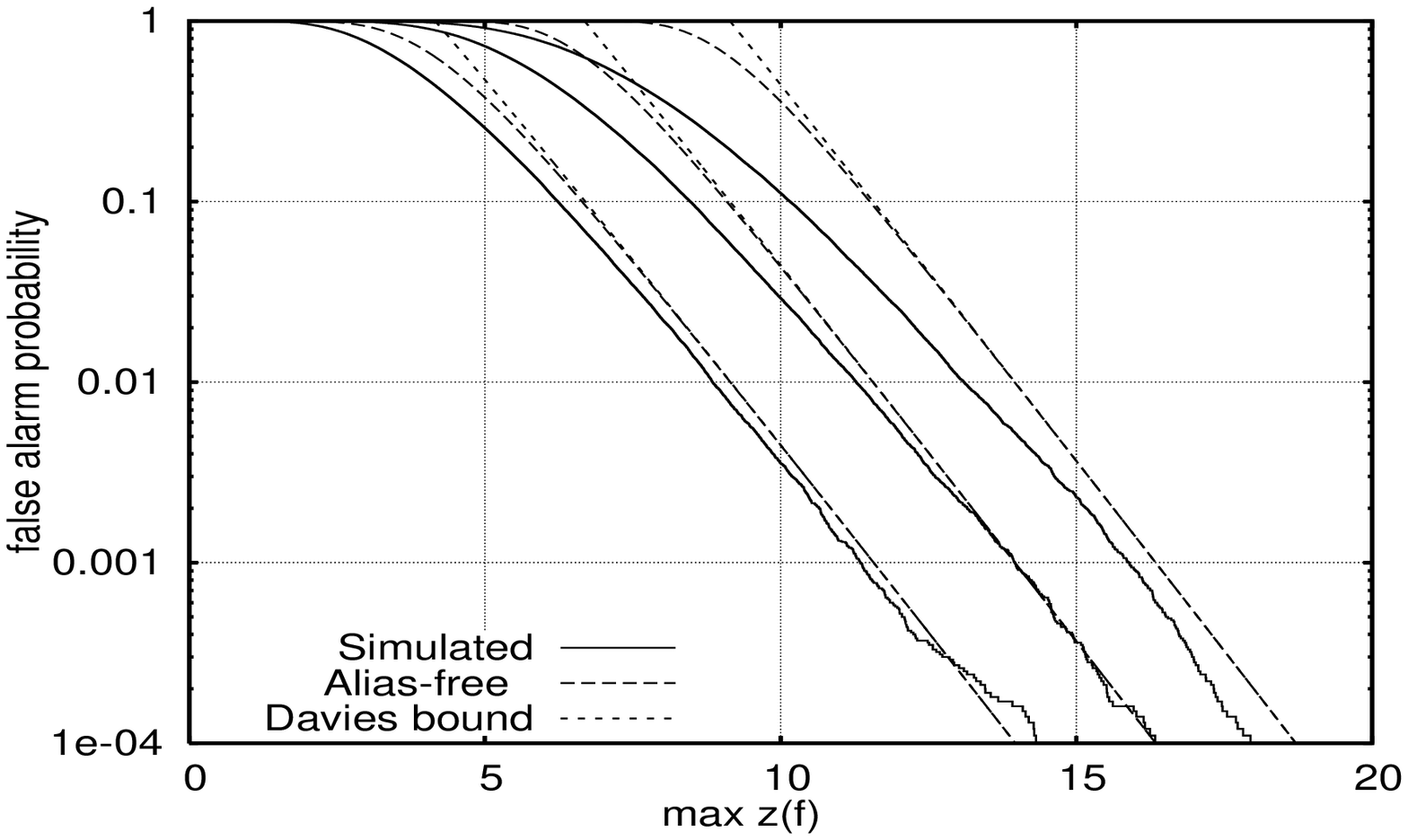}
\caption{The same as in Fig.~\ref{fig_simDistr51Peg}, but for the star 70~Vir.}
\label{fig_simDistr70Vir}
\end{figure}

The last pair of Monte-Carlo simulations in this paper deals with real astronomical time
series. I used epochs and standard errors of $153$ and $35$ radial velocity measurements
of the stars 51~Pegasi and 70~Virginis, obtained with ELODIE spectrograph
\citep{Naef04}\footnote{Note that both stars 51~Peg and 70~Vir have a planetary companion
\citep{MayorQueloz95,Marcy96}.}. These time series are not even. For the star 51~Peg, the
effective time series length $T_{\rm eff}\approx 9.0$~yrs is close to the actual one
$T\approx 9.2$~yrs, but the spectral window (Fig.~\ref{fig_simDistr51Peg}) shows several
high peaks indicating periodic gapping of observations. For the star 70~Vir, the time
series has $T_{\rm eff}\approx 8.5$~yrs, $T\approx 7.2$~yrs and posesses a more `noisy'
spectral window (Fig.~\ref{fig_simDistr70Vir}) indicating significant natural aliasing. In
the first case, the simulated extreme value distributions for the L--S periodogram don't
show large deviations from alias-free models (relative error $\Delta(\FAP)/\FAP
\lesssim 30\%$ and $\Delta z_*/z_* \lesssim 5\%$ for $\FAP<0.1$). In the second case, the
simulated $\FAP$ may be two times less than its alias-free approximation, but this still
may be tolerated because $\Delta z_*/z_* \lesssim 10\%$ (again for $\FAP<0.1$). Note that
the both time series possess a strong leakage with one day period. Such gapping affects
extreme value distributions for $P_{\rm min}=1/f_{\rm max}\leq 2$~days only. In the case
of 51~Peg, this aliasing could introduce a significant error in $\FAP$ for unrealistic
frequency ranges (say, for $P_{\rm min}=1/f_{\rm max}\lesssim 0.1$~days). In the case of
70~Vir, the respective deviation is enforced by low number of observations, what leads to
rather large errors of $\FAP$ already for $P_{\rm min}=1$~day. Note also that in the both
cases the errors of alias-free approximations decrease significantly when $\FAP$ drops to
the values $10^{-3}\div 10^{-2}$.

\begin{figure}
\includegraphics[width=84mm]{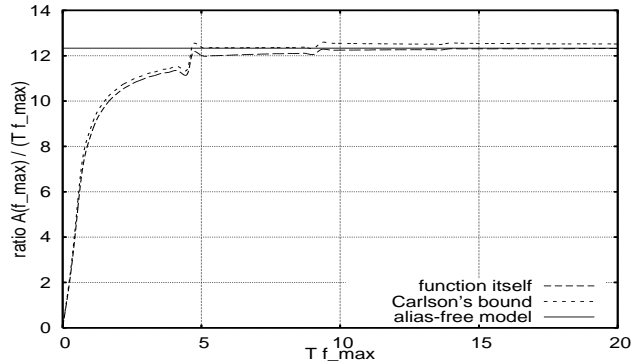}
\caption{The factor
$A(f_{\rm max})$ and its approximations.}
\label{fig_Lambda}
\end{figure}

At last, we need to consider the quality of the alias-free approximation for the factor
$A(f_{\rm max})$. Fig.~\ref{fig_Lambda} shows a graph of the ratio $A(f_{\rm max})/
(Tf_{\rm max})$ along with graphs of its alias-free approximation and upper Carlson bound
(see Appendix~\ref{sec_RiceSeries}). The observations were spanned in the same way as for
Fig.~\ref{fig_simDistrAlias}. The spectral leakage appears only in small splashes near the
Nyquist frequency of the periodic data breaks and near its overtones (i.e., at $f_{\rm
max}T= 4.5, 9.0, 13.5$). For $W>3$, the function $A(f_{\rm max})$ is well approximated by
the alias-free model regardless the strong aliasing.

\section{Conclusions}
The problem of estimating statistical significance of periodogram peaks is discussed in
the paper. The results published in the field of extreme values of random processes are
adapted for and extended to the periodogram analysis of astronomical time series. For the
Lomb--Scargle periodogram and its modifications the corresponding extreme value
distributions are given by a closed formulae being ready for usage. If the spectral
leakage cannot be neglected, the similar expressions provide upper limits to the false
alarm probability (or lower ones to the significance).

It is established numerically that the region of validity of these approximations is large
and has no sharp boundaries. Even if the aliasing is very strong, the error of the
analytic estimation of false alarm probability does not favour false alarms and thus is
not fatal. For strong aliases, the usage of this analytic approximation slightly decreases
the sensitivity to low-amplitude signals. However, the respective increasing of amplitude
thresholds should not exceed several per cent in the worst practical cases (like a strong
aliasing enforced by lack of observations).

These results may be very useful in a wide variety of astronomical applications. They
would be useful especially for systematic surveys that deal with large amounts of data
consisting of many separate time series. Indeed, it would be very difficult and even
impossible to perform Monte-Carlo simulation for every of such time series. Vice versa, it
is easy to use simple analytic formulae~(\ref{MaximaDistr},\ref{MaximaDistrIneq}) or their
analogs for the modified L--S periodograms. This will eliminate the need for Monte-Carlo
simulations in the cases when the observed periodogram peak exceeds the adopted threshold
and in the opposite cases when this peak is lower than this threshold by more than, say,
$10\%$. The rare intermediate cases are easy to be studied by means of Monte-Carlo
simulations. It is also admissible not to use numerical simulations at all, especially for
large datasets ($N\gtrsim 100$). In this case, the number of undetected low-amplitude
periodicities may be increased by a negligible quantity only.

\section*{Acknowledgments}
I would thank Drs. V.V.~Orlov, K.V.~Kholshevnikov, L.P.~Ossipkov and the anonymous referee
for critical reading of this paper, fruitful suggestions and linguistic corrections. This
work is supported by the Russian Foundation for Basic Research (Grants 05-02-17408,
06-02-16795) and by the President Grant NS-4929.2006.2 for the state support of leading
scientific schools.

\bibliographystyle{mn2e}
\bibliography{periodogram}

\appendix

\section{Several notations}
\label{sec_not}
Let us introduce the following averaging operations:
\[
 \langle \phi(t) \rangle = \sum_{i=1}^{N} \phi(t_i)/\sigma_i^2, \qquad
 \overline{\phi(t)} = \langle\phi(t)\rangle/\langle 1 \rangle, \nonumber
\]
with $\sigma_i^2$ being the error variance at the observational epoch $t_i$. The function
$\phi(t)$ may be defined at the set of $t_i$ only, that is to be a discrete sequence. The
quantity $\langle \phi_1(t) \phi_2(t) \rangle$ may be treated as a scalar product in the
Hilbert space \citep{SchwCzerny98a}.


All vectors are assumed to be column ones by default. The notation $\{x_1,x_2,\ldots\}$
corresponds to a column vector formed by the quantities inside the braces. Similarly,
$\{\bmath x_1,\bmath x_2,\ldots\}$ is a vector constituted by elements of the vectors
$\bmath x_1,\bmath x_2,\ldots$

$\Id$ is the identical matrix.

$*^T$ denotes the transpose of a matrix or a vector.

If $\bmath x$ is a vector then $\bmath x \otimes \bmath x := \bmath x \bmath x^T$ is a
matrix constituted by the pairwise products $x_i x_j$.


$p(x_1,x_2,\ldots)$ is the joint probability density of the random variables
$x_1,x_2,\ldots$ and $p(x_1=a_1,x_2=a_2,\ldots)$ is the same joint probability density,
taken in the point $(a_1,a_2,\ldots)$.


\section{Rice method and periodograms}
\label{sec_RiceSeries}
In the so-called `Rice method', one considers an integer random variable $N^+(Z_0,f_{\rm
max})$, the number of up-crossings of a given level $Z_0$ by the random process $Z(f)$
within $[0,f_{\rm max}]$. The distribution function of the maximum $Z_{\rm max} =
\max_{[0,f_{\rm max}]} Z(f)$ can be represented by the expansion
\begin{equation}
\Pr\{Z_{\rm max}\leq Z_0\} = \Pr\{Z(0)\leq Z_0\}\, \sum_{j=0}^\infty \frac{(-1)^j}{j!} \nu_j,
\label{RiceSeries}
\end{equation}
where $\nu_0=1$ and $\nu_j$ being the conditional factorial momenta of $N^+(Z_0,f_{\rm
max})$ under condition that $Z(0)\leq Z_0$. Let $\tilde\nu_j$ being the unconditional
factorial momenta of $N^+$. The quantities $\nu_j$ and $\tilde\nu_j$ are explicitly
expressed in terms of the so-called `Rice formulae'. For instance,
\begin{eqnarray}
 \tilde\nu_1(Z,f_{\rm max}) = \int\limits_0^{f_{\rm max}} df
 \int\limits_0^{\infty} Z' p(Z,Z') dZ', 
\label{Rice_1}\\
 \tilde\nu_j(Z,f_{\rm max}) =
 \int\limits_{[0,f_{\rm max}]^j} df_1\ldots df_j \int\limits_{[0,\infty)^j} Z'_1 \ldots Z'_j \times \nonumber\\
 \times\, p(Z_1=Z,Z'_1;\ldots; Z_j=Z,Z'_j)\, dZ'_1\ldots dZ'_j, 
\label{Rice_j}
\end{eqnarray}
where $p(Z,Z')$ being the joint probability density of $Z$ and $Z'=dZ/df$, both taken at
the same frequency $f$, and $p(Z_1,Z'_1;\ldots; Z_j,Z'_j)$ being the joint probability
density of the pairs $Z_i=Z(f_i),Z_i'=Z'(f_i)$. For details on the Rice method and further
references see the paper by \citet{Azais-RiceSeries}.

The expected number of up-crossings plays an important role in what follows. For the sake
of convenience, we introduce the synonymous notation $\tau \equiv \tilde\nu_1$. Exact
analytic expressions of $\tau(Z,f_{\rm max})$ for the periodograms $z(f)$ and $z_{1,2}(f)$
may be derived from results by \citet{Davies77,Davies87,Davies02}. Actually, Davies dealt
with the case when the weights of measurements are equal to each other. Nevertheless, his
results may be directly extended to the unequal weights. The quantity $\tau$ provides not
only the upper bound~(\ref{simFAP}) on the false alarm probability, but also yields its
asymptotic representation for large $z$ (low $\FAP$) levels. Unfortunately, the asymptotic
character of the Davies bound was strictly proved only for restricted families of random
processes, such as stationary Gaussian and stationary $\chi^2$ ones. Nevertheless, this
asymptotic seems to be non-specific to the distribution of the process values and to the
strict stationariness (see references and discussion in the cited works by Davies and in
the review by \citet{Kratz}). Hence, we may expect the asymptotic character
of~(\ref{simFAP}) for all of our periodograms. Note that the periodogram $2z(f)$ may be
treated as a $\chi^2$ random process, $2z_2/d$ as an $F$ process, and $2z_1/N_{\mathcal
H}$ as a beta process, according to \citet{Davies02}.

The high-order Rice formulae are significantly more complicated with respect to the
first-order one. We will not compute here the high-order Rice terms for our periodograms
in general case. However, the calculations are essentially simplified if the long-distance
correlations of the periodogram may be neglected (equivalently, the aliasing is
negligible). Indeed, under the approximation stated the density $p(Z_1,Z'_1;\ldots;
Z_j,Z'_j)$ may be factorized as $p(Z_1,Z'_1)\ldots p(Z_j,Z'_j)$ for all frequencies except
for the narrow vicinities of the diagonals $f_i= f_j$. This property allows us to obtain
that if $f_{\rm max}$ is large enough to be resolved by the periodogram well, the
relations $\nu_j\approx \tilde\nu_j \approx \tau^j$ hold true. Then the extreme value
distribution of $Z(f)$ is given by~(\ref{expo}). An alternative way to obtain the latter
expression is to assume a Poisson distribution for $N^+$ \citep{Kratz}.

The factor $A(f_{\rm max})$ in equalities~(\ref{tau_z},\ref{tau_z123}) determines the
dependence on the frequency range (so-called bandwidth penalty). Before considering it,
let us denote $\bvarphi_f'= \partial\bvarphi/\partial f$ and define the matrices
\begin{eqnarray}
 \mathbfss Q = \overline{\bvarphi \otimes \bvarphi}, \quad
 \mathbfss S = \overline{\bvarphi \otimes \bvarphi_f'}, \quad
 \mathbfss R = \overline{\bvarphi_f' \otimes \bvarphi_f'}, \nonumber\\
 \mathbfss Q_{\mathcal H} = \overline{\bvarphi_{\mathcal H} \otimes \bvarphi}, \qquad
 \mathbfss S_{\mathcal H} = \overline{\bvarphi_{\mathcal H} \otimes \bvarphi_f'}, \nonumber\\
 \mathbfss Q_{\mathcal H,\mathcal H} = \overline{\bvarphi_{\mathcal H} \otimes \bvarphi_{\mathcal H}}, \nonumber\\
 \tilde{\mathbfss Q} = \mathbfss Q - \mathbfss Q_{\mathcal H}^T \mathbfss Q_{\mathcal H,\mathcal H}^{-1} \mathbfss Q_{\mathcal H}, \qquad
 \tilde{\mathbfss S} = \mathbfss S - \mathbfss Q_{\mathcal H}^T \mathbfss Q_{\mathcal H,\mathcal H}^{-1} \mathbfss S_{\mathcal H}, \nonumber\\
 \tilde{\mathbfss R} = \mathbfss R - \mathbfss S_{\mathcal H}^T \mathbfss Q_{\mathcal H,\mathcal H}^{-1} \mathbfss S_{\mathcal H}, \qquad
 \mathbfss M = \tilde{\mathbfss Q}^{-1} (\tilde{\mathbfss R} - \tilde{\mathbfss S}^T \tilde{\mathbfss Q}^{-1} \tilde{\mathbfss S}).
\label{defM}
\end{eqnarray}
In general, all these matrices, except for the matrix $\mathbfss Q_{\mathcal H,\mathcal
H}$, depend on the frequency. Note the relations $\mathbfss S_{\mathcal H} = \mathbfss
Q_{\mathcal H}'$ and $\mathbfss S^T + \mathbfss S = \mathbfss Q'$. The
definitions~(\ref{defM}) look rather bulky, but they are essentially simplified under
certain conditions. For example, if the base functions $\bvarphi$ for any $f$ are
orthogonal to the functions $\bvarphi_{\mathcal H}$, then $\mathbfss Q_{\mathcal H} =
\mathbfss S_{\mathcal H} = 0$ and the matrices in~(\ref{defM}) labelled with a tilde are
equal to the same matrices without tilde mark. Also if the base $\bvarphi$ is orthonormal
for any $f$, then $\mathbfss Q = \Id$, $\mathbfss S^T = - \mathbfss S$ and $\mathbfss M =
\mathbfss R + \mathbfss S^2$. The last matrix $\mathbfss M(f)$ is necessarily positively
definite and possesses the positive eigenvalues $\lambda_i(f)$. In fact, we need below
only these eigenvalues. They satisfy the characteristic equation $\det(\tilde{\mathbfss R}
- \tilde{\mathbfss S}^T \tilde{\mathbfss Q}^{-1} \tilde{\mathbfss S} - \lambda
\tilde{\mathbfss Q}) = 0$.

The factor $A(f_{\rm max})$ appears implicitly in the papers by Davies after integration
of the mathematical expectation $\expect|\boldeta|$ by $f$, where the random vector
$\boldeta$ is Gaussian with zero mean and statistically independent componets having
$\disp\eta_i = \lambda_i(f)$. \citet{Davies87} gave some exact and approximate integral
formulae for this expectation. I present here (in terms of the factor $A$) a number of new
integral representations that may be useful in practice. The first two may be derived
easily and are given by
\begin{eqnarray}
 A(f_{\rm max}) &=& \int\limits_0^{f_{\rm max}} df \oint\limits_{\mathcal S_d}m(\bmath n)\, d\Omega,
                    \quad m^2(\bmath n) = \bmath n^T \mathbfss M \bmath n, \nonumber\\
 A(f_{\rm max}) &=& \int\limits_0^{f_{\rm max}} df \int\limits_{\bmath x^2<1}
                    \sqrt{\bmath x^T \mathbfss M \bmath x}\, \frac{d\bmath x}{|\bmath x|^d},
\label{coeffAinit}
\end{eqnarray}
where $d\Omega$ denotes an infinitesimal solid angle in $\mathbb R^d$, directed by the
unit-length integration vector $\bmath n$. The integration in the first formula is
performed over all possible directions within the whole space solid angle $S_d =
2\pi^{d/2}/ \Gamma(d/2)$. Note that the matrix $\mathbfss M$ may be diagonalized by means
of a solid-body rotation of $\bmath n$, so that $m^2 = \sum \lambda_i n_i^2$ and the
function $A(f_{\rm max})$ is determined by the eigenvalues $\lambda_i$ only. Inner
integrals in~(\ref{coeffAinit}) are equal to each other. They may be expressed in terms of
the (hyper)area of an ellipsoidal (hyper)surface in $d$ dimensions, having semi-axes $q_i=
\lambda_i^{-1/2}$. Indeed, changing the integration variable in the inner integral in the
second of equations~(\ref{coeffAinit}) as $\bmath x = \mathbfss M^{1/2} \tilde{\bmath x}$,
then $\tilde{\bmath x} = \tilde x \tilde{\bmath n}$ ($\tilde{\bmath n}^2=1$) and
integrating by $\tilde x$ we obtain
\begin{equation}
 \oint\limits_{\mathcal S_d} m(\bmath n) d\Omega = \Pi_d \sqrt{\det\mathbfss M}, \quad
   \Pi_d = \oint\limits_{\mathcal S_d} \frac{\sqrt{\tilde{\bmath n}^T \mathbfss M^2
   \tilde{\bmath n}}}{\left(\tilde{\bmath n}^T \mathbfss M \tilde{\bmath n}
   \right)^{\frac{d+1}{2}}}\, d\tilde\Omega.
\end{equation}
It can be directly checked that the integrand in the last expression represents an
infinitesimal (within $d\tilde\Omega$) area element on the surface $\tilde{\bmath x}^T
\mathbfss M \tilde{\bmath x} = \tilde x^2 \tilde{\bmath n}^T \mathbfss M \tilde{\bmath n}
= 1$, and $\Pi_d$ equals to its total area. It is not hard to show that $\Pi_1=2$. The
circumference of an ellipse, $\Pi_2$, and the usual surface area of an ellipsoid, $\Pi_3$,
can be expressed by means of elliptic integrals (complete and incomplete, respectively).
For $d\geq 4$ an Abelian integral can be used to compute $\Pi_d$ \citep{Tee-Ellipsoid}.
There are useful inequalities for $\Pi_d$, e.g. the Carlson's one bounds the inner
integrals in~(\ref{coeffAinit}) by the quantity $S_d\sqrt{(\lambda_1+ \ldots
+\lambda_d)/d} = S_d \sqrt{\trace\mathbfss M/d}$. Finally,
\begin{equation}
 A(f_{\rm max}) = \int\limits_0^{f_{\rm max}} \frac{\Pi_d(q_1\ldots q_d)}{q_1\ldots q_d} df
  \leq S_d \int\limits_0^{f_{\rm max}} \sqrt{\frac{\trace \mathbfss M}{d}}\, df.
\label{coeffA}
\end{equation}
The latter inequality seems to be very sharp in practical situations
(Fig.~\ref{fig_Lambda}). Note also, that if every $\lambda_i(f) \equiv \lambda$ then
$A(f_{\rm max})= S_d f_{\rm max} \sqrt\lambda$.

\bsp

\label{lastpage}

\end{document}